  \documentclass[
showpacs, amsmath, amssymb, floatfix,twocolumn,
nofootinbib,wrapfloat byrevtex]{revtex4}

\usepackage{minitoc}
\usepackage{epsfig}
\usepackage{amsmath}
\usepackage{mathrsfs}
\usepackage{graphicx}
\usepackage{bbm,verbatim}
\usepackage{color}
\usepackage{tikz}

\newcommand\tabcaption{\def\@captype{table}\caption}

\begin{document}
 \title{Thermodynamic fluctuation in black string flow}
 \author{Meng Sun$^{a}$\footnote{sunmeg.89@gmail.com}}
 \author{ Yong-Chang Huang$^{a,b}$ \footnote{ychuang@bjut.edu.cn}}
 \affiliation{ $^a$ Institute of Theoretical Physics\\ Beijing University of Technology, 100124, Beijing, China}
 \affiliation{ $^b$ CCAST(WorldLab.), P.O. Box 8730, 100080, Beijing, China}
 \begin{abstract}
 It has long been noticed that Laudau-Lifshitz theory can be used to study the fluctuation of a system that contains a black hole. Since the black string can be constructed by extending n-dimensional black hole into one extra dimension. We study the fluctuation of black string flow with a Schwarzschlid-like metric in D=n+1 dimensional spacetime and a charged solution in D=5 dimensional spacetime and get the second moments of the fluctuation of the mass flux and charge flux.
 
 \end{abstract}
 \maketitle

 \tableofcontents

\section{Introduction}

String theory is a competitive candidate for the quantum theory of gravity and the unification of all interactions in nature.
Because string theory requires extra dimensions to keep its consistency, it gives us a strong motivation to study the higher dimensional black hole when we consider gravity in string theory.

Recently, an exact solution for a black string flow was given in \cite{1}. 
In such a configuration, we do not need a negative cosmological constant to assure the existence of the black string which is necessary for the black string motivated by AdS/CFT correspondence \cite{2,3,4}.
In this construction the horizon of this black string flow is closely similar to black funnels in \cite{2,3,4,9,10,11,12}.

The construction of the black string flow is as flows. A thin black string falls freely into a very large black hole. The size of the black string $r_{bs}$ and black hole $r_{bh}$ have such a relation $r_{bh}\gg r_{bs}$. Since the black string is free-falling, one can expect that the two horizons merge into each other smoothly.
In this picture, because the black string is falling, the thin black string will pour mass (and charge) to the large black hole. As a consequence, the black hole and black string can be regarded as a non-equilibrium system. Since we have the $r_{bh}\gg r_{bs}$, it is reasonable to say that the current of the flow is relatively small. If we study this phenomenon in a small time interval, which means that the time scale of any measurement we do here is much shorter than the time scale of the process itself, we can study the non-equilibrium thermodynamic fluctuations of the growing black hole that forms on the infinitely large black hole `background' i.e. the contact part. 
The method we used here is the Laudau-Lifshitz theory \cite{5} which has been successfully used to study the evaporation of different kinds of black holes\cite{6,7,8,Wang:1995mw,D9,D10}. By this theory, we found the second moment of fluctuation of different black string flow and what kind of parameters can affect the fluctuation.

This article is organised as: In section 2, we study the evolution of horizon flow. In section 3, we use the Laudau-Lifshitz theory to analyse the fluctuation of the flow. In section 4, we introduce charges into the system to see how these will affect the fluctuation. Section 5 is the conclusion and outlook.

\section{Null hypersurface and its evolution}

Suppose that we have a thin black string whose radium is $r_{hs}$, and freely falls into a infinitely large black hole. To illustrate the whole process, we plot a general graph. In Fig.1, the black line denotes the black string, the gray part denotes the infinitely large black hole, the dashed line is the place where the black string and black hole meet and the arrows point the direction of the black string flow.
\begin{figure}
\centering
\begin{tikzpicture}[xscale=13,yscale=3.8]
	\draw [very thick, <->] (0,0.8)--(0,0)--(0.51,0);
	\node at (0,0.85) {z};
	\node at (0.53,0) {r};
	\path [fill= gray] (0.0015,0.2) rectangle (0.5,0.0055);
	\draw [very thick, -> ] (0.05,0.8)--(0.05,0.5);
	\draw [very thick] (0.05,0.5)--(0.05,0.3);
	\draw [thick, ->] (0,0.6)--(0.05,0.6);
	\node at (0.02,0.65) {$r_{bs}$};
	\draw [dashed, very thick] (0.05,0.3) to [out=-90, in=180] (0.1,0.2);
	\draw [dashed, very thick, ->] (0.1,0.2)--(0.25,0.2);
	\draw [dashed, very thick] (0.25,0.2)--(0.45,0.2);
\end{tikzpicture}
\caption{\label{fig:1}An illustration of the black string flow. The black line denotes the black string, the gray part denotes the infinitely large black hole, the dashed line is the place where the black string and black hole meet and the arrows point the direction of the black string flow.}
\end{figure}
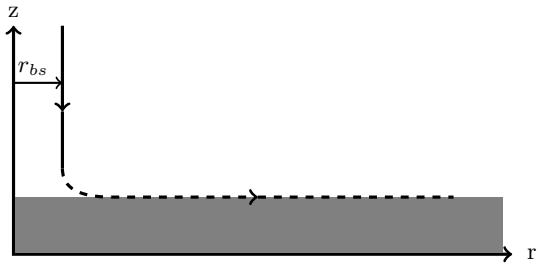 

In $D=n+1$ spacetime, the black string has the general form
\begin{equation}
	ds^2=g_{\mu \nu}dx^\mu dx^\nu +dz^2
\end{equation}
where $g_{\mu\nu}dx^\mu dx^\nu$ can be the metric of any $n$ dimensional black hole spacetime.
When $g_{\mu \nu}$ is Schwarzschild,
the resulting metric is
\begin{equation}\label{1}
ds^2=-\left(1-\left(\frac{r_0}{r}\right)^n\right) dt^2+dz^2+\frac{dr^2}{1-\left(\frac{r_0}{r}\right)^n}+r^2 d\Omega_{n+1}
\end{equation}
When the black string is free falling, based on \cite{1}, we get
\begin{equation}\label{7}
\frac{dr}{dt}=\left(\frac{r_0}{r}\right)^{\frac{n}{2}}-\left(\frac{r_0}{r}\right)^{\frac{3n}{2}}
\end{equation}
,
\begin{equation}\label{8}
\frac{dz}{dt}=1-\left(\frac{r_0}{r}\right)^n
\end{equation}
and the horizon of the black string flow.
In Fig.2, we plot the behaviour of Eq.\eqref{8} with $n=2$ and $r_0=5$. The derivative of $r$ with respect to $t$ decreases as the radius increases. When the radius is large enough the variation of $r$ is relatively small comparing to the size of the flowing system.  
In Eq.\eqref{8}, the derivative of $z$ with respect to $t$ is the speed of the falling black string. When $r$ is large enough this speed can be regarded as a constant. In Fig.3, we show the appearance of horizon of the black string flow and the motion of the horizon when  $n=2$ and $r_0=1$ where the curves are the outline of the horizon at $t=5$ and $t=7$. In Fig.4, we show the evolution of the horizons on time when  $n=2$ and $r_0=1$. This shows the outline of the horizon at different time. When $r$ is small the growing is significant, but when $r$ is large this phenomenon is insignificant i.e. the process is very slow. Since when $r$ is very large the speed of the flow is a steady flow and the growing of the horizon is very slow, we can use the Laudau-Lifshitz theory to study this system.  
\begin{figure}
\centering
\includegraphics[width=2in,clip]{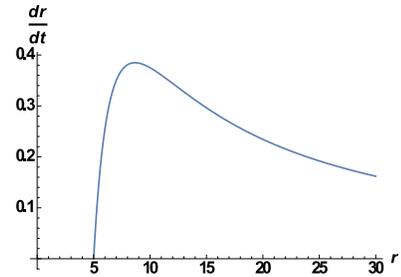}
\caption{\label{fig:3} The behaviour of the Eq.\eqref{7} when $n=2$ and $r_0=5$. As the r increases, the change of the r decreases significantly.}
\end{figure}

\begin{figure}
\centering
\includegraphics[clip, bb= 0 0 310 210, scale=0.6]{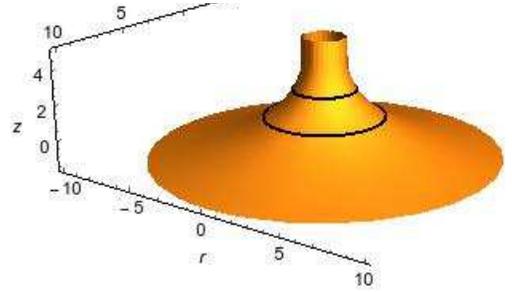}
\caption{\label{fig:4} The appearance of horizon of black string flow when  $n=2$ and $r_0=1$. The curves show the outline of the horizon at different time.}
\end{figure}
\begin{figure}
\centering
\includegraphics[width=2in,clip]{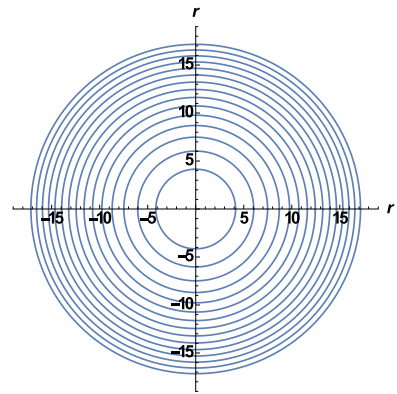}
\caption{\label{fig:5}The outline of the horizon at different time in $r$ direction. The first circle in center is the outline of the horizon at $t=10$. The last circle is the outline of the horizon at $t=150$. The time interval between every two neighbouring circles is the same, $dt=10$.}
\end{figure}
Now we take a slice of the black string at the point where the black string and the black hole meet together. The segment can be regarded as a black hole. Recalling that the n-dimensional volume of a Euclidean ball of radius R in n-dimensional Euclidean space is
\begin{equation}\label{9}
V_n\left(R\right)=\frac{\pi^{\frac{n}{2}}}{\Gamma\left(\frac{n}{2}+1\right)}R^n
\end{equation}
where $\Gamma$ is gamma function. The surface area of n-sphere of radius R is the boundary of the n+1-ball of radius R. The we have the following relation
\begin{equation}\label{10}
A_n\left(R\right)=\frac{d}{dR}V_{n+1}\left(R\right)
\end{equation}
Then the surface area of the black hole formed on the segment is
\begin{eqnarray}\label{11}
A_{n+1}\left(r_h\right)&=&\frac{d}{dr_h}V_{n+2}\left(r_h\right)\\
&=&\left(n+2\right)\frac{\pi^{\frac{n}{2}+1}}{\Gamma\left(\frac{n}{2}+2\right)}\left( r_h^{n+1}-r_0^{n+1}\right)\nonumber
\end{eqnarray}
In Eq.\eqref{11} we subtract the $r_0^{n+1}$ term, because it is the surface area of the black string.
Then the entropy of the black hole in $D=n+4$ spacetime dimension is
\begin{equation}\label{12}
S_{n+4}=\frac{A_{n+1}}{4}=\left(n+2\right)\frac{\pi^{\frac{n}{2}+1}}{\Gamma\left(\frac{n}{2}+2\right)}\left( r_h^{n+1}-r_0^{n+1}\right)
\end{equation}
The $r_h$ in Eqs. \eqref{11} and \eqref{12} is the horizon of the black hole which is formed by the black string flow. Then we have
\begin{equation}\label{13}
\frac{dS_{n+4}}{dt}=\left(n+1\right)\left(n+2\right)\frac{\pi^{\frac{n}{2}+1}}{\Gamma\left(\frac{n}{2}+2\right)}r_h^{n} \frac{dr_h}{dt}
\end{equation}
This is the evolution of the entropy with respect to time.

\section{Non-equilibrium thermodynamic fluctuation}
When the size of the black string $r_0$ is much smaller than the size of the black hole at the point of their contact, we may approximate the process to a non-equilibrium steady state process. We can use Landau-Lifshitz theory to study the non-equilibrium thermodynamic fluctuations of the black string flow. Such a theory has been applied well in the black hole system\cite{6,7,8,Wang:1995mw,D9,D10}.

According to the Landau-Lifshitz theory, in a fluctuation-dissipative process the flux $\dot{x}$ for a given thermodynamic quantity $x_i$ is
\begin{equation}\label{14}
\dot{x}_i=-\Gamma_{ij}X_j
\end{equation}
and the rate of entropy production is
\begin{equation}\label{15}
\dot{S}=\pm X_i\dot{x}_i
\end{equation}
Here the positive sign represents the contribution to the entropy rate from the non-concave parts in the entropy and the minus sign is for other conditions. Then the second moments in the fluctuation of the fluxes is\cite{5}
\begin{equation}\label{16}
\langle\delta \dot{x}_i \delta \dot{x}_j\rangle= \left(\Gamma_{ij}+\Gamma_{ji}\right)\delta_{ij}
\end{equation}
From \eqref{14} to \eqref{16}, the quantities $\Gamma_{ij}$ are the phenomenological transport coefficients (or Onsager phenomenological transport coefficients) and $X$ is the thermodynamic force that is conjugate to the flux $\dot{x}_i$. The dot represents their derivative with respect to time t.

For a given length in $z$ direction, $z=L$, the total mass of the black string is
\begin{equation}
M=Lm_0
\end{equation}
where $r_0=2m_0$.
By substituting it in \eqref{8}, we have
\begin{equation}\label{17}
\dot{m}=m_0\left(1-\left(\frac{2m_0}{r}\right)^n\right)
\end{equation}
Since we have \eqref{13}, we can get
\begin{widetext}
\begin{eqnarray}\label{18}
\dot{S}_{n+4}&=&\frac{d}{dt}\left(\left(n+2\right)\frac{\pi^{\frac{n}{2}+1}}{\Gamma\left(\frac{n}{2}+2\right)}\left( r^{n+1}-r_0^{n+1}\right)\right)
=2^{n+1}\left(n+2\right)\left(n+1\right)\frac{\pi^{\frac{n}{2}+1}}{\Gamma\left(\frac{n}{2}+2\right)}m_0\left(1-\left(\frac{2m_0}{r}\right)^n\right)m^n\nonumber
\end{eqnarray}
\end{widetext}
Then according to Eq.\eqref{15} the thermodynamic force conjugate to $\dot{m}$ is
\begin{equation}\label{19}
F_M=-2^{n+1}\left(n+2\right)\left(n+1\right)\frac{\pi^{\frac{n}{2}+1}m^n}{\Gamma\left(\frac{n}{2}+2\right)}
\end{equation}
What we have calculated above is not the non-concave part of the entropy.
According to Eq.\eqref{14}, the transport coefficient with respect to this is
\begin{equation}\label{20}
\Gamma_{mm}=\frac{m_0\left(1-\left(\frac{2m_0}{r}\right)^n \right)\Gamma \left(\frac{n}{2}+2\right)}{2^{n+1} \left(n+2\right) \left(n+1\right)\pi^{\frac{n}{2}+1}m^n}
\end{equation}
Then by Eq.\eqref{16}, we have the second moments in the fluctuation of the mass flux
\begin{equation}\label{21}
\left\langle \delta \dot{m} \delta \dot{m} \right\rangle = \frac{m_0\left(1-\left(\frac{2m_0}{r}\right)^n \right)\Gamma \left(\frac{n}{2}+2\right)}{2^{n} \left(n+2\right) \left(n+1\right)\pi^{\frac{n}{2}+1}m^n}
\end{equation}
In Eqs.\eqref{18}-\eqref{21}, $m$ denotes the mass which has already flowed on the black hole from the black string.
Then based on Eqs.\eqref{17} and \eqref{18} the second moment in the fluctuation of entropy is
\begin{widetext}
\begin{equation}\label{37}
\left\langle \delta \dot{S}_{n+4} \delta \dot{S}_{n+4} \right\rangle= 2^{n+1}\left(n+2\right)\left(n+1\right)\frac{\pi^{\frac{n}{2}+1}}{\Gamma\left(\frac{n}{2}+2\right)}m^n \left\langle \delta \dot{m} \delta \dot{m} \right\rangle
\end{equation}
\end{widetext}
The second moment in the fluctuation of entropy shows that it is driven by two kinds of parameters: the first one is the parameter related with the black hole that is $m$; the second one is the parameter related with the string flow that is $r$. In Eq.\eqref{37}, the fluctuation of entropy can not diverge.
\section{Fluctuation for charged flows}
For a more general black string metric we can introduce the electric charges. Let $m_0$ and $q_0$ denote the mass and charge of the black string and let $m$ and $q$ denote the mass and charge already on the black hole. For simplicity we just consider the condition that the dimension is D=5
\begin{equation}\label{22}
ds^2=-\Delta dt^2  +\frac{dr^2}{\Delta}+\mathcal{Z}dz^2+r^2d\Omega^2
\end{equation}
with
\begin{equation}\label{23}
\Delta=1-\frac{2m_0}{r}+\frac{q_0^2}{r^2}
\end{equation}
where $\mathcal{Z}$ is a metric function which should be positive outside the black sting horizon and should tend to 1 at $r\rightarrow\infty$.
Comparing with the previous analysis \cite{1}, we have
\begin{equation}\label{25}
\frac{dr}{dt}=\Delta\sqrt{\frac{\mathcal{Z}-\Delta}{\Delta}} ~ ~ and ~ ~ \frac{dz}{dt}=\frac{\Delta}{\mathcal{Z}}
\end{equation}
Under the D=5 condition, the entropy of the charged black hole formed by the flowing charged black string is
\begin{equation}\label{26}
S_5=\frac{A_2}{4}=\pi \left(R_+^2 - r^2_+\right)
\end{equation}
Here $r_+=m_0+\sqrt{m_0^2-q_0^2}$ is the outer horizon of the black string and the $R_+=m+\sqrt{m^2-q^2}$ is the outer horizon of black hole. According to Eq.\eqref{26}, the derivative of the entropy with respect to time is
\begin{equation}\label{27}
\dot{S}=2\pi \frac{R_+}{m^2-q^2}\left(R_+\dot{m}-q\dot{q}\right)
\end{equation}
Based on Eq.\eqref{25}, the mass flux is
\begin{equation}\label{28}
\dot{m}=m_0\frac{\Delta}{\mathcal{Z}}+\frac{q}{R_+}\dot{q}
\end{equation}
and the charge flux is
\begin{equation}\label{29}
\dot{q}=q_0\frac{\Delta}{\mathcal{Z}}
\end{equation}
In Eq.\eqref{28}, the second term accounts for the coupling between the gain rates of the mass and electric charge\cite{13}.
These give us the second moments in the fluctuation of the flux(for details, see appendix A)
\begin{equation}\label{30}
\left\langle \delta \dot{m} \delta \dot{m} \right\rangle = \frac{\left(m^2-q^2\right) m_0 \Delta}{\pi R_+^2\mathcal{Z}}
\end{equation}
\begin{equation}\label{31}
\left\langle \delta \dot{q} \delta \dot{q} \right\rangle = -\frac{2\left(m^2-q^2\right)q_0\Delta}{\pi  q R_+\mathcal{Z}}
\end{equation}
and
\begin{equation}\label{32}
\left\langle \delta \dot{m} \delta \dot{q} \right\rangle = -\frac{\left(m^2-q^2\right)q_0\Delta }{\pi R_+^2\mathcal{Z}}
\end{equation}
The second moments in the entropy can be calculated from Eq.\eqref{27}, we can get
\begin{widetext}
\begin{equation}\label{33}
\left\langle \delta \dot{S} \delta \dot{S} \right\rangle = \left(2\pi \frac{R_+}{m^2-q^2}\right)^2 \left( R_+^2 \left\langle \delta \dot{m} \delta \dot{m} \right\rangle +q^2 \left\langle \delta \dot{q} \delta \dot{q} \right\rangle -2q R_+ \left\langle \delta \dot{m} \delta \dot{q} \right\rangle \right)
\end{equation}
\end{widetext}
According to statistical mechanics, the phase transitions are usually related to the divergence of the relevant second moments\cite{5}. Obviously, when $m=q$ the Eq.\eqref{33} diverges. This indicates that a critical phenomenon takes place. This phenomenon is only controlled by the variable of the black hole formed by the black string flow. However, when $\Delta=\mathcal{Z}$ Eqs.\eqref{30}-\eqref{32} show that the second moments in the fluctuation of the fluxes are independent of the metric of the black string flow. In Ref.\cite{1}, this indicates that the system reaches a thermal equilibrium state. In this state, $\Delta=\mathcal{Z}$, there is no flow actually. Eqs. \eqref{30}-\eqref{32} then just show fluctuations around the thermal equilibrium state. In addition, we can define such a parameter $\eta=m-q$. Then the second moments Eq.\eqref{33} diverge when $\eta \rightarrow 0$. Basing on the arguments in \cite{8}, $\eta$ can be regarded as an order parameter in the Landau order-disorder phase transition. The $\eta=0$ case represents the symmetric phase, in which the whole system is under the extremal condition. The $\eta\neq0$ represents the symmetry broken phase, in which the whole system is under non-extremal condition.
\section{Conclusion and outlook}
The black string flow we analyse here is an approximate system that a black string smoothly merges into a black hole. In this paper, we investigated the non-equilibrium thermodynamic fluctuations of the black string flow and charged black string flow. We have considered black hole growing as a non-equilibrium thermodynamic process because the black string pours mass (and charge) into the black hole. According to Landau-Lifshitz theory, the divergence of second order momentum of thermodynamic fluctuation is the signature of phase transitions, we find that in black string flow the second momentum of the flux does not diverge. However, in charged black string flow condition, the second momentum of the flux is divergent at $m=q$, i.e. at the extremal conditions.  Even though the phase transition does not depend on the property of the black string, the parameters such as $\Delta$ and $\mathcal{Z}$ of the black string can affect the fluctuation of the black hole. When the system reaches thermodynamic equilibrium state at $\Delta=\mathcal{Z}$, the fluctuation is free from the black string, the phase transition only depends on the parameter of the black hole. 
Such a fluctuation may affect some observations at astrophysical scale, for example, through the gravitational lensing\cite{Pal:2007ap} and the motion of test particles\cite{Abdujabbarov:2009az}\cite{Stuchlik:2008fy}.

A similar configuration, the black string with rotation (and charged)\cite{14}, can be used to study the fluctuation in that system. In section 4, we just considered the charged string that is the black string with electric charges. For the string-charged, i.e., the strings are electric sources of a 2-form potential $B_{\mu\nu}$, or 0-brane charge that sources a Maxwell potential $A_\mu$. Since the black string flow solution with charge and rotation can reach a thermodynamic equilibrium at the extreme limit, we may study the thermodynamic equilibrium fluctuation of the black string flow by using the thermodynamic curvature that has been applied well in this field\cite{15,16,17,18,19,20,21,22}.
\section*{Acknowledgments}
We gratefully acknowledge discussions with Dr. Ding-Fang Zeng, Nan Li and all the others in our institute. The work is supported by National Natural Science Foundation of China (No. 11275017 and No. 11173028).\\

\appendix
\begin{Large}
\textbf{Appendix}
\end{Large}

\section{Calculations about the fluctuation of charged flows }
In this appendix, we give the details about the calculations of Eq.\eqref{30} in section 4.

According to Eqs.\eqref{14} and \eqref{28}, we have
\begin{equation}\label{34}
m_0\frac{\Delta}{\mathcal{Z}}+\frac{q}{R_+}\dot{q}=-\Gamma_{mm}F_M-\Gamma_{mq}F_Q
\end{equation}
In Eq.\eqref{28}, the first term relates to the mass flow directly, the second term is the coupling between the gain rates of the mass and the electric charge. This is due to a fact that the potential energy of black string flow will increase when this flow pours down to the black hole.
Then we can write Eq.\eqref{34} into two independent terms. They are
\begin{equation}\label{35}
m_0\frac{\Delta}{\mathcal{Z}}=-\Gamma_{mm}F_M ~ ~ and ~ ~ \frac{q}{R_+}\dot{q}=-\Gamma_{mq}F_Q
\end{equation}
Then basing on Eqs.\eqref{15} and \eqref{27}, we know the thermodynamic forces conjugate to the flux of mass and charge are\footnote{In Eq.\eqref{15} $X_i$ is a general expression. Here $F_M$ and $F_Q$ are the forces conjugate to the flux of mass and charge.}
\begin{equation}\label{36}
F_M=-2\pi\frac{R_+^2}{m^2-q^2}~ ~ and ~ ~ F_Q=2\pi\frac{R_+q}{m^2-q^2}
\end{equation}
According to Eq.\eqref{35} the transport coefficient are
\begin{equation}
\Gamma_{mm}=\frac{m_0 \Delta\left( m^2-q^2\right)}{2\pi\mathcal{Z}R_+^2} ~ ~ and ~ ~ \Gamma_{mq}=-
\frac{\left(m^2-q^2\right)q_0 \Delta}{2\pi R_+^2 \mathcal{Z}}
\end{equation}
Then we can get Eq.\eqref{30}, and given the Onsager relation $\Gamma_{ij}=\Gamma_{ji}$ we have Eq.\eqref{32}. We can use the similar process to get Eq.\eqref{31}, too.


\begin{thebibliography}{99}
  \bibitem{1} R. Emparan, M. Martinez,  \emph{JHEP} {\bf 09} (2013) 068 [arXiv:1307.2276]
	\bibitem{2}
		S. Fischett, D. Marolf,  \emph{Class.\ Quant.\ Grav.\ }{\bf 29} (2012) 105004 [arXiv:1202.5069]
	\bibitem{3}
		P. Figueras, T. Wiseman, \emph{Phys.\ Rev.\ Lett.\ }{\bf 110} (2013) 171602 [arXiv:1212.4498]
	\bibitem{4}
		S. Fischetti, D. Marolf and J.E. Santos,  \emph{Class.\ Quant.\ Grav.\ }{\bf 30} (2013) 075001 [arXiv:1212.4820]
		\bibitem{9}
		V.E. Hubeny, D. Marolf and M.Rangamani,  \emph{Class. Quant. Grav.} {\bf 27} (2010) 025001 [arXiv:0909.0005]
	\bibitem{10}
		V.E. Hubeny, D. Marolf and M.Rangamani,  \emph{Class. Quant. Grav.} {\bf 27} (2010) 095018 [arXiv:0911.4144]
	\bibitem{11}
		M.M. Caldarelli, O.J. Dias, R. Monteiro and J.E. Santons, \emph{JHEP} {\bf 05} (2011) 116 [arXiv:1102.4337]
	\bibitem{12}
	  	J.E. Santos and B. Way, \emph{JHEP} {\bf 12} (2012) 060 [arXiv:1208.6291]
	\bibitem{5}
		L. D. Landau, E. M. Lifshitz, \emph{Statical Physics}
	\bibitem{6}
		D. Pav\'on, and Rubi, J.M., \emph{Phys. Rev. D }  {\bf 37}, 2052 (1988)
	\bibitem{7}
	 	D. Pav\'on , \emph{Phys. Rev. D } {\bf 43}, 2495 (1991)
	\bibitem{8}
		Su, R.K. Cai. R.G. and Yu. P.K.N. , \emph{Phys. Rev. D } {\bf 50}, 2932 (1994)
		\bibitem{Wang:1995mw} 
  B.~Wang and J.~M.~Zhu,
  Mod.\ Phys.\ Lett.\ A {\bf 10}, 1269 (1995).
	\bibitem{D9}
D. Pavón and J. M. Rubrndoti aai, Phys. Lett. 99A, 214 (1983).
	\bibitem{D10}
D. Pavón and J. M. Rubrndoti aai, Gen. Relativ. Gravit. 17, 387 (1985)		
	 \bibitem{13}
		W.A. Hiscock and L.D. Weems, \emph{Phys. Rev. D.} {\bf 41} 1142 (1990)
		\bibitem{Pal:2007ap} 
  S.~Pal and S.~Kar,
  Class.\ Quant.\ Grav.\  {\bf 25}, 045003 (2008)
  [arXiv:0707.0223 [gr-qc]].
  \bibitem{Abdujabbarov:2009az} 
  A.~Abdujabbarov and B.~Ahmedov,
  Phys.\ Rev.\ D {\bf 81}, 044022 (2010)
  [arXiv:0905.2730 [gr-qc]].
\bibitem{Stuchlik:2008fy} 
  Z.~Stuchlik and A.~Kotrlova,
  Gen.\ Rel.\ Grav.\  {\bf 41}, 1305 (2009)
  [arXiv:0812.5066 [astro-ph]].

	\bibitem{14}
		M. Sun and Y.C. Huang  [arXiv:1405,6906]
	\bibitem{15}
	J.~E.~Aman, I.~Bengtsson and N.~Pidokrajt,
  Gen.\ Rel.\ Grav.\  {\bf 35}, 1733 (2003)
  [gr-qc/0304015].
  \bibitem{16}
  G.~Ruppeiner,
  Phys.\ Rev.\ D {\bf 78}, 024016 (2008)
  [arXiv:0802.1326 [gr-qc]].
  \bibitem{17}
   G.~Arcioni and E.~Lozano-Tellechea,
  Phys.\ Rev.\ D {\bf 72}, 104021 (2005)
  [hep-th/0412118].
	\bibitem{18}
	R.~G.~Cai and J.~H.~Cho,
  Phys.\ Rev.\ D {\bf 60}, 067502 (1999)
  [hep-th/9803261].
  \bibitem{19}
  J.~y.~Shen, R.~G.~Cai, B.~Wang and R.~K.~Su,
  Int.\ J.\ Mod.\ Phys.\ A {\bf 22}, 11 (2007)
  [gr-qc/0512035].
  \bibitem{20}
   C.~Niu, Y.~Tian and X.~N.~Wu,
  Phys.\ Rev.\ D {\bf 85}, 024017 (2012)
  [arXiv:1104.3066 [hep-th]].
	\bibitem{21}
	R.~Banerjee, S.~Ghosh and D.~Roychowdhury,
  Phys.\ Lett.\ B {\bf 696}, 156 (2011)
  [arXiv:1008.2644 [gr-qc]].
  	\bibitem{22}
  	Y.~D.~Tsai, X.~N.~Wu and Y.~Yang,
  Phys.\ Rev.\ D {\bf 85}, 044005 (2012)
  [arXiv:1104.0502 [hep-th]].


\end{thebibliography}
 \end{document}